# Simulation of electron behavior in PIG ion source for 9MeV cyclotron

X. J. Mu[1], M. Ghergherehchi[1], Y.H. Yeon[1], J. W. Kim[1], J.S. Chai[1a]

[1]College of Electronic and Electrical Engineering, Sungkyunkwan University, Suwon, Korea

**Abstract** In this paper, we focus on a PIG source for producing intense H-ions inside a 9 MeV cyclotron. The properties of the PIG ion source were simulated for a variety of electric field distributions and magnetic field strengths using CST particle studio. After analyzing secondary electron emission (SEE) as a function of both magnetic and electric field strengths, we found that for the modeled PIG geometry a magnetic field strength of 0.2 T provided the best results in terms of the number of secondary electrons. Furthermore, at 0.2 T the number of secondary electrons proved to be greatest regardless of the cathode potential. Also the modified PIG ion source with quartz insulation tubes was tested in KIRAMS-13 cyclotron by varying gas flow rate and arc current, respectively. The capacity of the designed ion source was also demonstrated by producing plasma inside the constructed 9MeV cyclotron. As a result, the ion source is verified to be capable of producing intense H⁻ beam and high ion beam current for the desired 9 MeV cyclotron. The simulation results provide experimental constraints for optimizing the strength of the plasma and final ion beam current at target inside a cyclotron.

---

[a] jschai@skku.edu

## 1. Introduction

Ion sources are used for producing intense ion beams in various accelerators. The PIG ion source is well known as an internal ion source inside a cyclotron. It derives its name from the vacuum gauge invented by Penning [1].

Computer simulation of charged particle beam is an important tool for scrutinizing processes occurring in various fields of physics [2, 3]. Multiply computer simulations by different codes in ion source have been done before, and the simulation results are useful for identification and improvement of ion source device. Simulation of H$^-$ ion source extraction systems is performed using ion beam simulator IBSIMU developed at the University of Jyvaskyla. The simulation results and experimental data were in good agreement [4, 5]. Other researchers have also demonstrated good experimental and simulation agreement when considering an improved PIG ion source with hydrogen particles within the KIRAMS-13 cyclotron. Thus, the simulation results repeatedly confirm that optimizing the ion source structure and applied field strengths plays a critical role for good operation of the cyclotron [6].

Therefore, in this paper, we used CST PARTICLE STUDIO (CST PS) [7] to do simulation and analysis of electron behavior in a PIG ion source for a 9 MeV cyclotron. The objective of the PIG ion source simulation within CST PS is to obtain constrains on the electric potential at the cathode and the magnetic field to gain high density plasma. We note that while visualizing the electron trajectories and determining SEE characteristics is one part of the ion source operation, a comprehensive analysis requires inclusion of electron interaction effect with the target gas, which was not down here. Igniting and sustaining dense plasma requires that enough high energy electrons present in the system to replenish those lost from interacting with the neutral gas. As part of the simulations we varied the magnetic field and electric field to try to optimize the production of secondary electrons, finding an optimum magnetic field strength of 0.2 T. The KIRAMS-13 cyclotron which produces 13MeV energy ion beam was developed by Korea Institute of Radiological Medical Sciences (KIRAMS) to produce short-lived positron emission tomography radioisotopes for medical diagnosis and treatment. It is used at more than 5 regional cyclotron centers in South Korea. The maximum produced beam current is limited to the expected value, due to low efficiency in ion source performance. Multiply investigations to evaluate the ion beam current were studied and performed in previous work [6], [8-11].

## 2. Technical parameters between 13 MeV and 9 MeV cyclotrons

The main parameters of the 13 MeV cyclotron and 9 MeV cyclotron are illustrated and compared in Table 1. While the shape of magnet inside a cyclotron is an essential consideration that could affect beam acceleration, however the PIG ion source here is unaffected by these changes.

To compare any possible effect on the PIG ion source designed in this study, here we compare the essential parameters that determine the ion source ionization process for the 9 MeV and 13 MeV cyclotron systems. Both of the vacuum pressures inside these two cyclotrons are provided by diffusion pumps which are capable of making $10^{-7}$Torr pressure in 1 hour. The magnetic field distributed at the central region provides homogenous magnetic field for ionization process inside the PIG ion source is approximately the same value (1.99T, 1.89T). Dee voltage connected to the central region provides the extraction voltage on puller to extract H- beam from the ion source is 40 kV and 45 kV. Therefore, by demonstrating the ion beam extraction capabilities from the aperture on the anode in the 13 MeV system, we have shown that the source can successfully work in other systems with different acceleration voltages, such as the desired 9 MeV cyclotron.

Table 1. Main parameters of the 13 MeV and 9 MeV PET cyclotrons.

| Parameter | 13MeV cyclotron | 9MeV cyclotron | Unit |
|---|---|---|---|
| Maximum energy | 13 | 9 | MeV |
| Beam species | Negative hydrogen | Negative hydrogen | - |
| Ion source | PIG/ Internal | PIG/ Internal | - |
| Maximum B field | 1.99 | 1.89 | Tesla |
| Central B field | 1.288 | 1.36 | Tesla |
| Vacuum pumps | Diffusion pumps | Diffusion pumps | - |
| Radio-frequency | 71.5 | 83 | MHz |
| Dee voltage | 40 | 45 | Kilo volts |
| Harmonic number | 4 | 4 | - |
| Ion extraction | Stripper foil | Stripper foil | - |
| Cooling system | Water cooling | Water cooling | - |

3. **Simulation Procedure in CST PS**

CST PS is a software tool for 3-dimension design and investigation of charged particle trajectories in the presence of an electromagnetic field [8]. In the PIG ion source operation, the emission of secondary electrons is an important parameter determining the behavior and efficiency of plasma generation. There are three basic types of secondary electrons: backscattered, re-diffused secondary and elastic reflected electrons [12]. We note that CST PS utilizes a model by Furman and Pivi [13], and is a well-known simulation environment for analyzing the generation of secondary electrons.

The model of PIG ion source was drawn based on the geometry of a typical PIG ion source. The geometry of the model and intensity of the electric and magnetic fields were defined firstly, along with other necessary components, including cathode, anode, cathode holder, pocket and cover. The cathodes were 7 mm in diameter and 2.0 mm in thickness, and were screwed into the cathode holders. The cylinder shape anode with 20 mm in length having different internal diameters of 6.16 mm to 10.1 mm, were used in simulation.

The anode was selected as SEE copper and the cathodes were chosen as tantalum material. Note that due to material parameter definition, the secondary electron emission only occurs on the inside surface of the anode, although in a real experimental setting emission it will also occur at the cathode/anticathodes. The pocket, cathode holds and cover parts were set as aluminum, which is non-magnetic. Further, all conducting materials were defined as perfect electrically conducting (PEC) while using vacuum as the background material corresponding to an experimental environment.

As all the components for the electrostatic part were defined, we opted to simulate the electric potential from 100 V up to 2500 V, while considering the magnetic field strengths of 0.2, 0.4, and 0.6 T. These values were chosen due to experimental lab constraints, and also due to previous work for a separate (but similar) geometry in the literature [14]. To apply a magnetic field in a PIG, either a permanent magnet or an electromagnetic coil can be utilized experimentally.

For CST, an electromagnetic coil implantation was utilized, with magnetic fields defined by the number of coil turns N and electric current I, based on Eq (1).

$$NI = \oint \frac{B}{\mu} dl = B \times \frac{l_1}{\mu_0} + B \times \frac{l_2}{\mu} \approx B \times \frac{l_1}{\mu_0} \quad (1)$$

where:

$l_1$ = the length of gap between two cathodes,

$\mu_0$ = the permeability in air,

B = magnetic field intensity, (B=0.2 T in experimental permanent magnet)

The initial seed electrons were set up after definition of the electric and magnetic fields. The seed electrons were then set to fixed emission from the cathode surfaces as shown in Fig. 1 (purple dots). Particle trajectories were calculated within the particle track simulation workflow with a time-step defined by the fastest particle in the simulation (i.e. the time steps will not be constant).

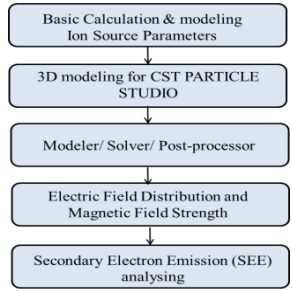

The simulation setup model

## 4. Results and discussion

### 4.1 Electron trajectory

Firstly, we demonstrate visualization of the electron trajectories in -1000 volt electric field and 0.1T density magnetic field for the PIG ion source in Fig. 2a and 2b respectively. The magnetic field is homogeneous distributed, and two fields are almost paralleled to each other.

Prior to considering the effect of magnetic field and electric field on the formation of secondary electrons, we examined the effect of an off-center cathode positioning with respect to the anode and magnetic coil. As could be expected, the properly aligned configuration produces significantly more secondary emission electrons because the diameter of electron trajectory is bigger as seen in Figure 3. It was also observed that the lifetime of the secondary electrons was also longer, due to better confinement of the electrons.

Next the effect of varying the electrical field, while holding the magnetic field strength constant, was considered. High density electrons which can make many electron-neutral collisions happen are required for efficient ionization of a neutral gas. Further, electron energy should be several times the magnitude of the neutral atom ionization potential. Since directly this condition is difficult to achieve, an abundance of secondary electrons must be emitted to reach a higher energy than the primary electrons. To illustrate the dependence of SEE on cathode potential, Fig. 4 shows the number of secondary electrons (Each simulation has the same density/number of particle sources from cathode, thus the number of secondary electrons is comparable) for voltages -100 V to -2500 V at the cathode, and for B = 0.2, 0.4, and 0.6 T. The graph clearly shows that at B = 0.2T, the greatest number of secondary electrons are produced. However, as a function of electric potential, the 0.4T magnetic field produces a more consistent/stable number of secondary electrons. As the magnetic field is further increased, the SEE counts drop considerably, as can be seen for 0.6 T. We also plot the secondary electron kinetic energy as a function of the applied cathode potential at 0.2 T in Fig.5. It's observed that the kinetic energy is increased as the value of electric field increases.

Therefore, for the geometry considered in this paper, a suggested magnetic field strength for SEE is 0.2 T. This result and the simulation data, while not suggestive of an optimal plasma-generation condition, can be utilized for future simulations including the effect of neutral gas ionization effects.

*4.2 Anode geometry in CST PS*

The anode was designed as illustrated in Fig. 6. To verify its advantage, this designed anode and the prototype in cylinder shape were modeled in this work using CST PS. The number of secondary electrons in these two anodes varied by different density of electric fields is demonstrated in Fig.7 three times more secondary electrons are produced by using the new designed anode compared with the cylinder one. Consequently, the designed anode which can produce bigger amount of higher energy secondary electron is proved to produce higher density of plasma since it makes higher ionization probability.

Herein the evaluation of secondary electron emission by varying the inner diameter of anode is also simulated using CST PS. The prototype anode is 6.16 mm in inner diameter. The electron distribution is calculated as a function of anode inner diameter while placed in a 2000 volt electric field and 1.36 Tesla magnetic field, as illustrated by Fig. 8 for fixed value of emitted electron. As a result in this work, the anode with 7 mm in inner diameter is demonstrated to be capable of producing the highest density of electrons while the 8 mm inner diameter anode gives the highest beam current density in KIRAMS-13 [8]. Fig.8 also shows that when the anode with inner diameters higher than 9 mm, the number of electron production will decrease, this means the reduction of secondary electron production efficiency.

*4.2 Ion beam extraction in CST PS*

In order to optimize the beam emission from the central region to the acceleration region, ion beam extraction from ion source side wall aperture through typical puller was modeled using CST PS. The H- beam was extracted under the following experimental constraints based on the KIRAMS-13 cyclotron: -2.5 kV voltage on cathodes by the arc power supply, 45 kV voltage on pullers from Dee voltage, a 1.288T center homogeneous magnetic field [6]. As a simulation result, the relative location of pullers to ion source is fixed at the distance of 2.2 mm between the center of two pullers and the anode aperture in horizontal direction. By rotating the anode, it is found that the ratio of beam extraction through puller is a distribution over the anode rotation angles. The ion beam extracted from slit on anode wall through puller enters the acceleration region inside the cyclotron, and part of the beam is lost because of beam diffusion flow in the central region (Fig. 9). It is demonstrated from Fig. 10 that with the anode rotation angles from -0.2 to -1.5 degree, more than 50% H⁻ beam can be extracted through pullers. The H⁻beam extraction trajectory is shown in Fig. 9. The simulation result is useful for optimizing future beam extraction from puller in the 9 MeV cyclotron.

*4.3 Test and measurement of modified PIG ion source in KIRAMS-13 cyclotron*

The ion source with new designed insulator tubes was tested and measured in a KIRAMS-13 cyclotron. The relative angle between the ion source and puller was fixed at 30 degrees while the relative angle between the slit and puller was fixed at 1.5 degrees based on the CST simulation results. Beam currents at the carbon foil and the target according to gas flow rate and arc current were measured. The results are shown in Figs.11-12, respectively. It is verified that the beam current at carbon foil or target is increased when both the gas flow rate and the arc current are increased. To investigate the influence of the two factors in the ion source performance, one factor was fixed while the other one was varied. As illustrated in Fig. 12, the arc current induces a weaker effect (by percentage change) on the beam current at both the carbons foil and the target when compared with simply varying the gas flow rate. The maximum beam current measured at target was 48.3 micro amperes under 2.0 ampere arc current and 7.5 sccm hydrogen gas flow rate.

## 5. Conclusion

We optimized the experimental facilities using both simulations and experiments. Computer simulation of electron trajectory in the PIG ion source for 9 MeV cyclotron was successfully implemented in CST PARTICLE STUDIO. The dependence of SEE on cathode potential and confining magnetic field strength were also examined, with the greatest SEE occurring for 0.2 T. The result is helpful for future simulation and experiment to generate high density of plasma and high ion beam current. The behavior of electrons, modification of ion source geometry and optimization of H$^-$ beam extraction were modeled in CST PARTICLE STUDIO. The simulation results provide experimental constraints for optimizing the plasma strength and beam current inside a cyclotron. The modified ion source was tested at different gas flow rates and arc currents in the KIRAMS-13 cyclotron, and it was demonstrated gas flow rate and arc current play important roles in the H$^-$ beam current. Moreover, the ion beam extraction capabilities from the aperture on the anode wall were demonstrated. We note a PIG ion source in a 9 MeV cyclotron at Sungkyunkwan University constructed from the same design was recently shown to successfully produce a beam.

## ACKNOWLEDGEMENT

This work was supported by the IT R&D program of MKE KEIT. [10043897 Development of a 500 cGy level radiation therapy system based on automatic detection and tracing technology with dual-head gantry for 30% reduction in treatment time for cancer].

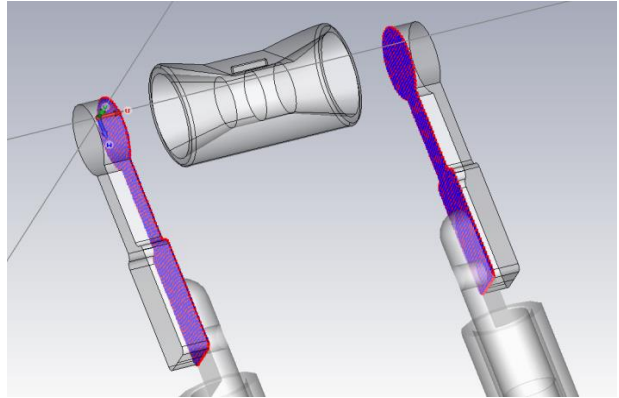

**Fig. 1.** Particle source from cathode surfaces

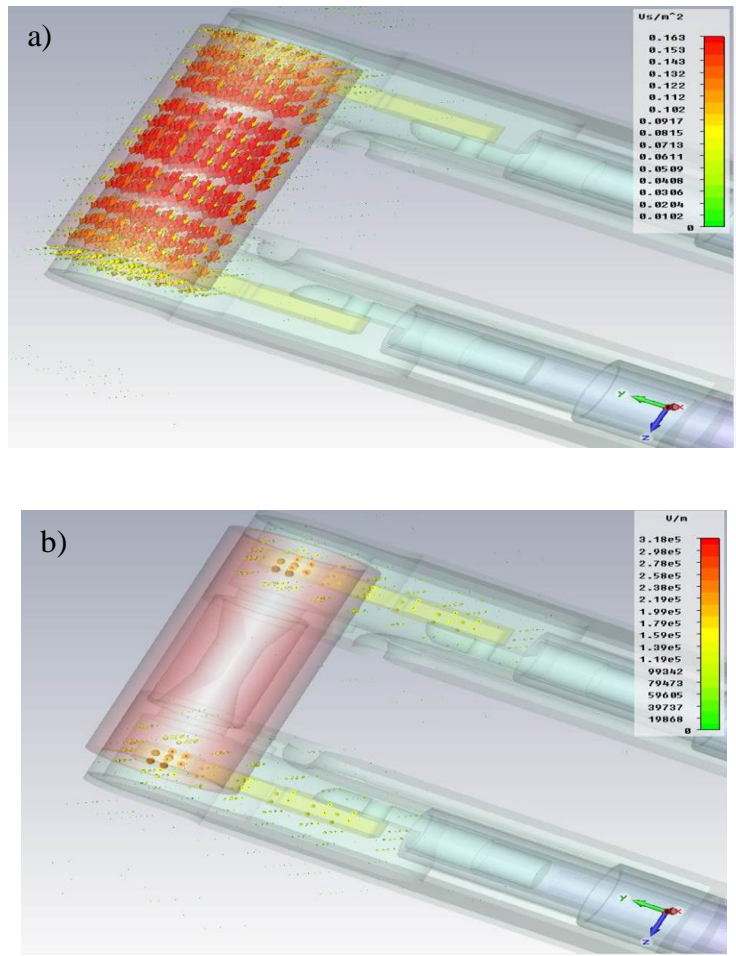

**Fig. 2.** The electron trajectories plotted in a) 0.1T magnetic field and b) -1000V electric field

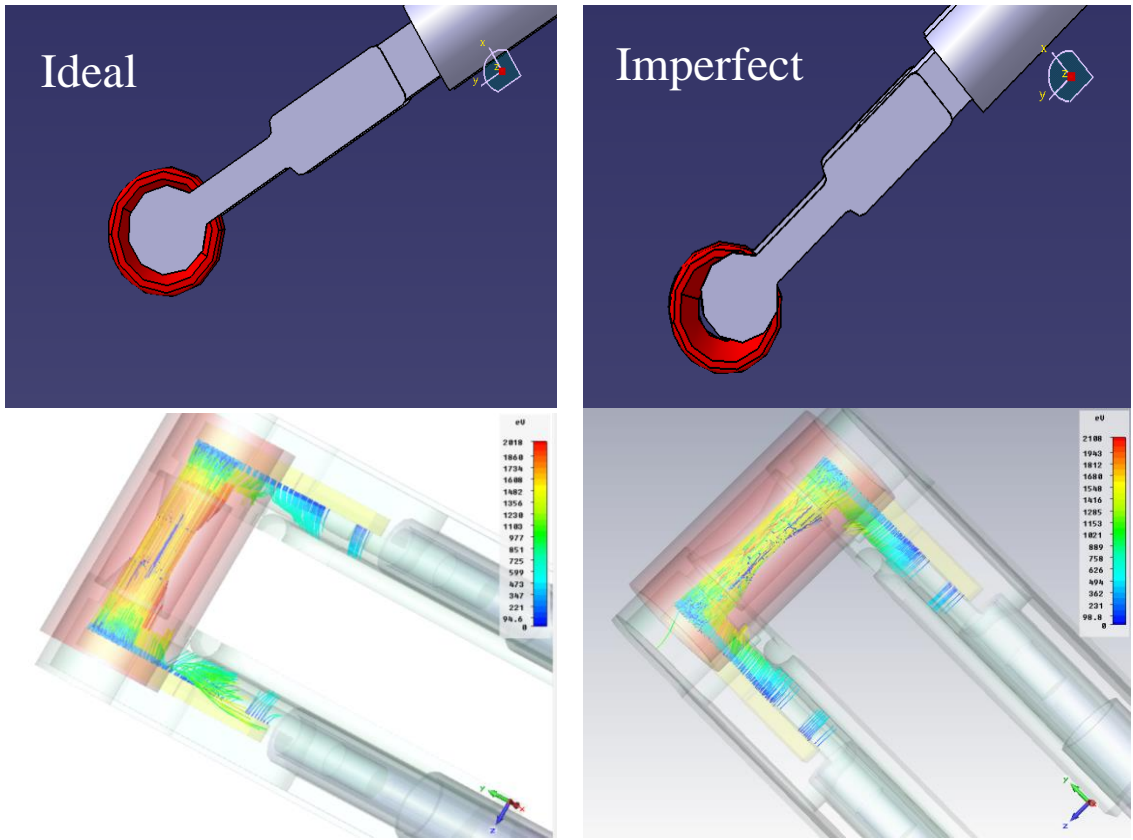

**Fig. 3.** Geometry and electron trajectories corresponded of ideal and imperfect geometry

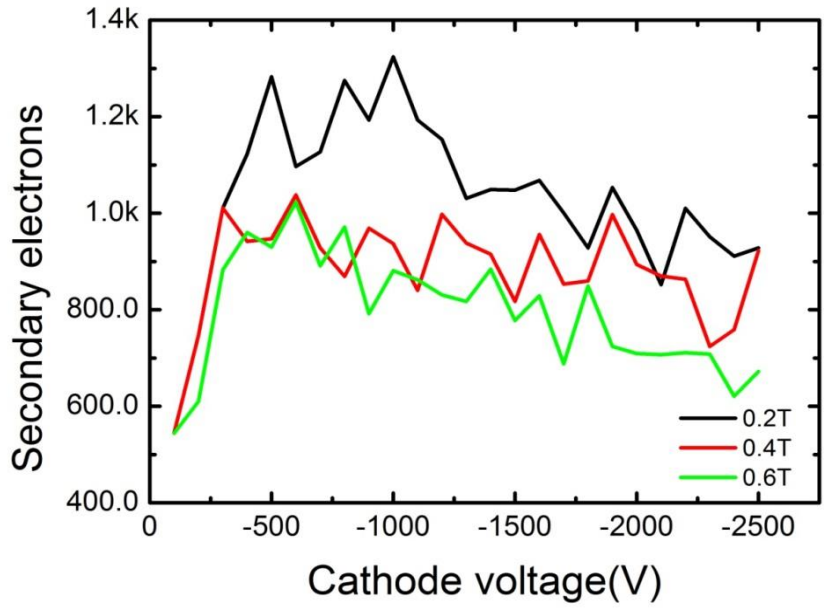

**Fig. 4.** Voltage applied to cathode vs number of secondary electron in 0.2T, 0.4T, 0.6T magnetic field.

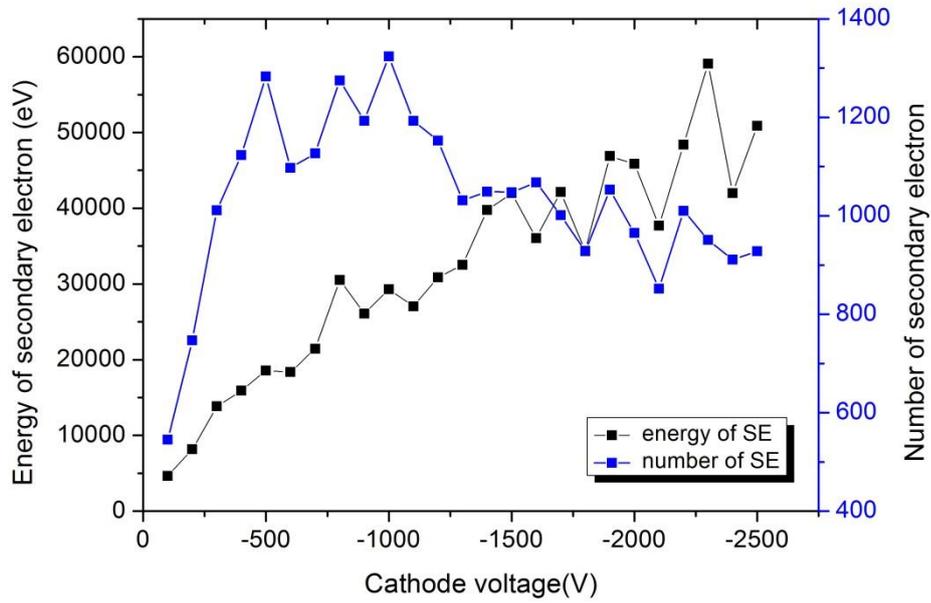

**Fig. 5.** Energy of secondary electrons in 0.2T magnetic field.

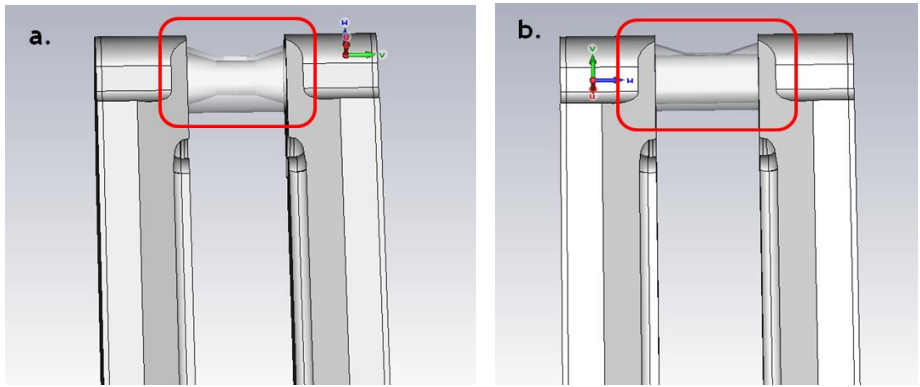

**Fig. 6.** PIG ion source with a). Designed anode; b). Inner cylinder shape anode.

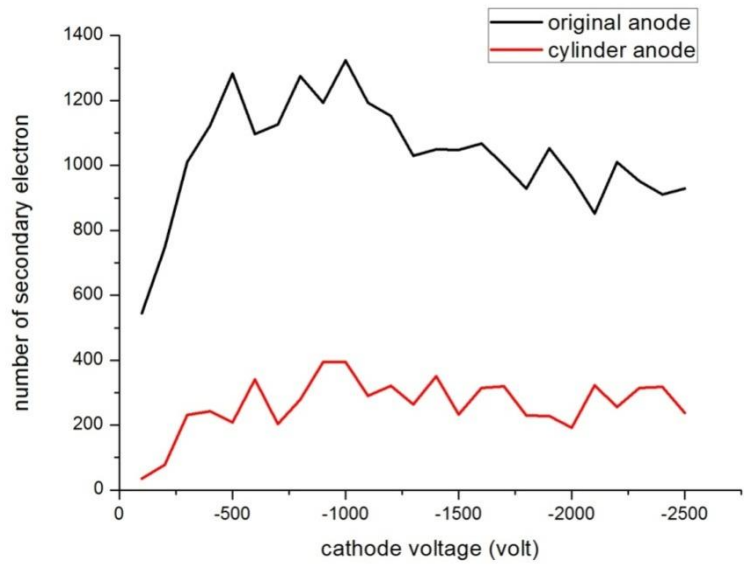

**Fig. 7.** Number of secondary electrons vs. cathode voltages in two types of anodes

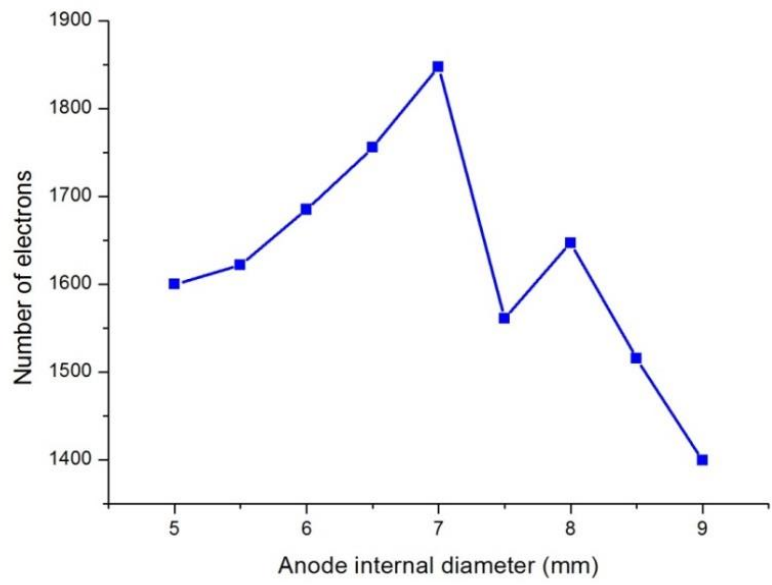

**Fig. 8.** Simulation results of anode with various inner diameters.

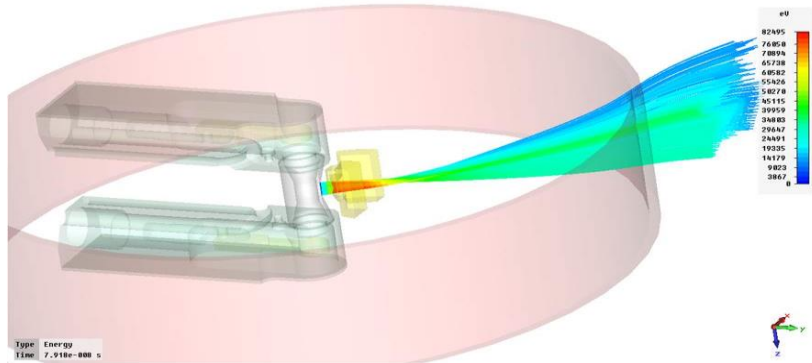

**Fig. 9.** Extracted ion trajectory through pullers.

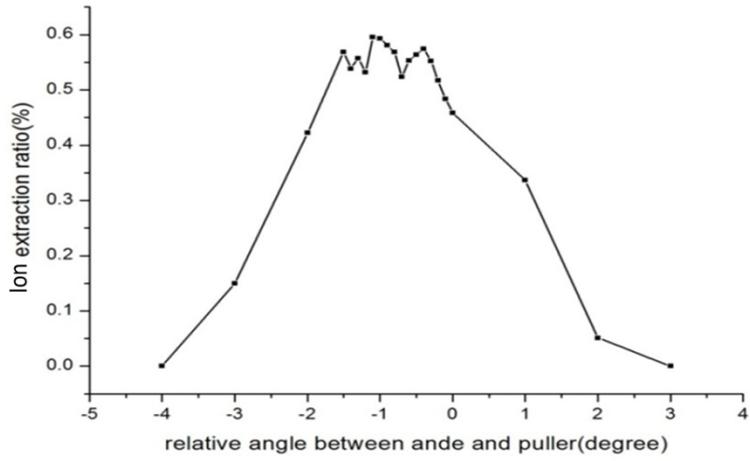

**Fig. 10.** Ion extraction ratio vs. anode rotation angles

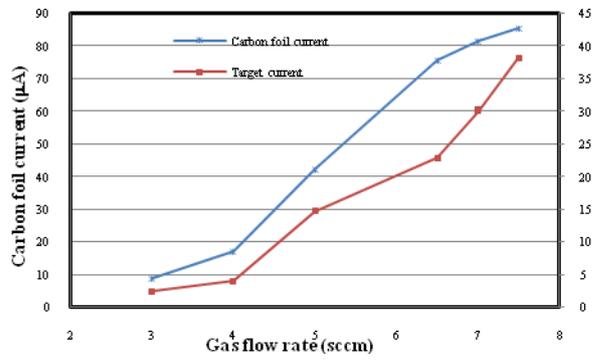

**Fig. 11.** Beam current at carbon foil and target vs. gas flow rate.

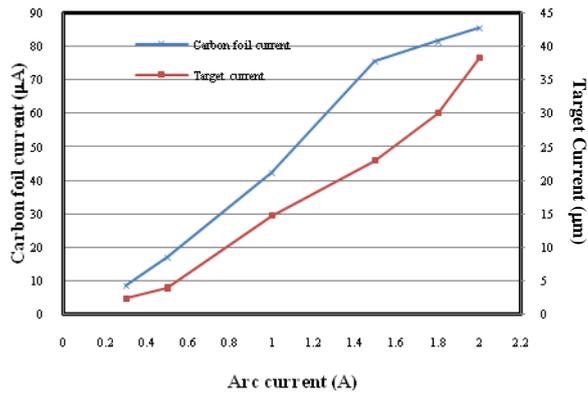

**Fig. 12.** Beam current at carbon foil and target vs. arc current.